\documentclass[pre,aps,twocolumn,showpacs]{revtex4}
\usepackage{graphicx}
\usepackage{amssymb,amsfonts,amsmath}

\begin{document}

\title{Generic Criticality in Ecological and Neuronal Networks}
\author
{David A. Kessler}
\noindent
\affiliation{Department of Physics, Bar-Ilan University,
 Ramat-Gan, IL52900 ISRAEL}
 \author
{Herbert Levine}
\noindent
\affiliation{Center for Theoretical Biological Physics, Rice University, Houston TX 77096, USA}

\begin{abstract}
We investigate the dynamics of two models of biological networks with purely suppressive interactions between the units; species interacting via niche competition and neurons via inhibitory synaptic coupling. In both of these cases, power-law scaling of the density of states with probability  arises without any fine-tuning of the model parameters. These results argue against the increasingly popular notion that non-equilibrium living systems operate at special critical points, driven by there by evolution so as to enable adaptive processing of input data.
\end{abstract}
\maketitle

Are biological systems adjusted by evolution to operate at a special critical point?\cite{bialek1} This idea has surfaced in variety of different contexts over the past years, including network architectures~\cite{kauffman,banavar}, scaling in neural processing~\cite{chialvo, bialek1,bialek2}, and self-organization of biological flocks~\cite{parisi,bialek3}. Underlying this concept is the appealing but vague notion that being critical is somehow conducive to flexible adaptation in the face of large variations in stimuli to be acted upon by the system in question.

One approach to demonstrating this special critical nature relies upon a maximum entropy approach to fit correlations of the fundamental events~\cite{bialek1,bialek4}, say spikes in the neural context. This method constructs an effective Hamiltonian for the system that reproduces the equal-time correlations and expectation values. This Hamiltonian then allows for the evaluation of the specific heat as a function of temperature $T$, where by definition $T=1$ is used to fit the original data. A specific heat peak at $T=1$ which grows as a function of system size is then interpreted as evidence of criticality of the actual biological process.

Here, we explore the extent to which this picture is valid, using two different classes of biological models arising respectively in the ecological and neural contexts. These models have in common that the nonlinear interaction between the fundamental entities are purely inhibitory. Species inhibit each other's survival  by competing for the same niche~\cite{may} and also our neurons have purely inhibitory couplings as they attempt to classify incoming activation signals. In both of these cases, non-trivial power-law scaling of the integrated density of states ${\cal{N}}(p)$ with the probability $p$ emerges without having to carefully adjust parameters or consider sophisticated feedback scenarios. In the language of Ref. \cite{bialek4}, the entropy $S \equiv log {\cal{N}}(p)$ is linear in the energy. Our systems are inherently non-equilibrium, with a power-law distribution of net fluxes (see later) and hence cannot be described by any Hamiltonian. Forcing an equilibrium Hamiltonian to reproduce the expectation values and equal-time correlations leads to a theory that looks critical in the aforementioned sense, but this feature is not a useful characterization of the actual operating state of these processes. 

\begin{figure}
 \centering{\includegraphics[width=0.45\textwidth]{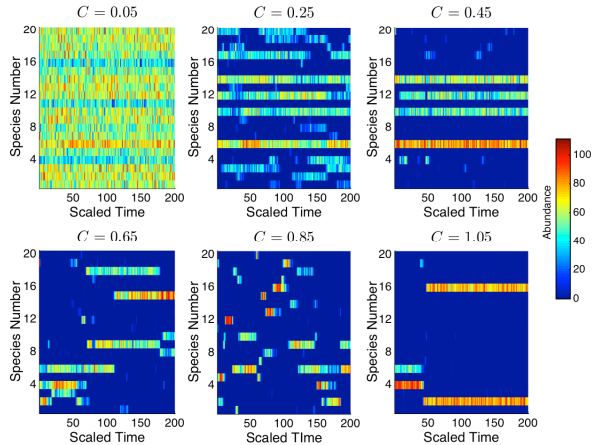}}
 \caption{Snapshots of the abundance for all species in the stochastic model for various levels of competition, $C$.   Red is high abundance, dark blue low, as shown in the
color bar at the right.  $Q=20$, $K=100$,
 $\sigma=0.5$, $\lambda=0.01$. The time is measured in units of  $0.08/\lambda$.  The first panel exhibits full coexistence, the second and third partial coexistence, the fourth and fifth disorder dynamics, and the last glass-like dynamics. (From Ref. \cite{kessler1})}
 \label{figGLVsnaps}
 \end{figure}

{\bf Generalized Competitive Lotka-Volterra System} We start by reviewing a recent study~\cite{kessler1} of a generalized Lotka-Volterra model~\cite{volterra} to study competing ecological species~\cite{may,banavar2}. This model describes a semi-isolated island with infrequent immigration of individuals from an external ``mainland" on which reside $Q$ species.  The individuals on the island undergo a stochastic birth-death process, where the competition-induced death rate, $d_i$, for individuals of a given species $i$ is determined by the abundances, ${n_k}, k=1,\ldots Q$ of all the various species (including its own) on the island,
\begin{equation}
d_i = \frac{1}{K}\left(n_i +  C\sum_{j\ne i} c_{i,j} n_j\right) ,
\end{equation}
where $K$ represents the carrying capacity and $C$ measures the strength of competition. The interaction strengths $c_{i,j}$ are chosen randomly from a Gamma distribution with mean unity and standard deviation $\sigma$.
 For simplicity, the birth rates of all the different species are chosen to be identical, a value which is set equal to unity to define the time scale.  The immigration rates of the various species are also fixed to be identical, and  denoted by $\lambda$.  This model was studied intensively~\cite{kessler1}, and found to exhibit four different regimes, or ``phases", of behavior as the competition strength $C$ is varied. For small $C$, the competition effects are minor and typically all $Q$ species are present on the island in steady-state (the full coexistence phase). At some critical $C$, the weakest species basically goes extinct, except for occasional ultimately unsuccessful attempts to recolonize the island via immigration. Increasing $C$ further causes more species to undergo extinction (the partial coexistence phase). At a second critical $C$, the dynamics changes dramatically, and there is no longer any set of species which is permanently represented.  Rather, the system undergoes regime shifts where, due to the combined effects of extinction and immigration, the set of dominant species changes completely from time to time (the disordered phase).  Lastly, at $C$'s close to unity, the system is marked by long periods where some small set of species coexists, with infrequent regime shifts between a few such sets (the glass-like phase).  This behavior is illustrated in Fig. \ref{figGLVsnaps}, where for various $C$'s, snapshots of the abundance of all $Q=20$ species through time is presented.
 
\begin{figure}
 \centering{\includegraphics[width=0.45\textwidth]{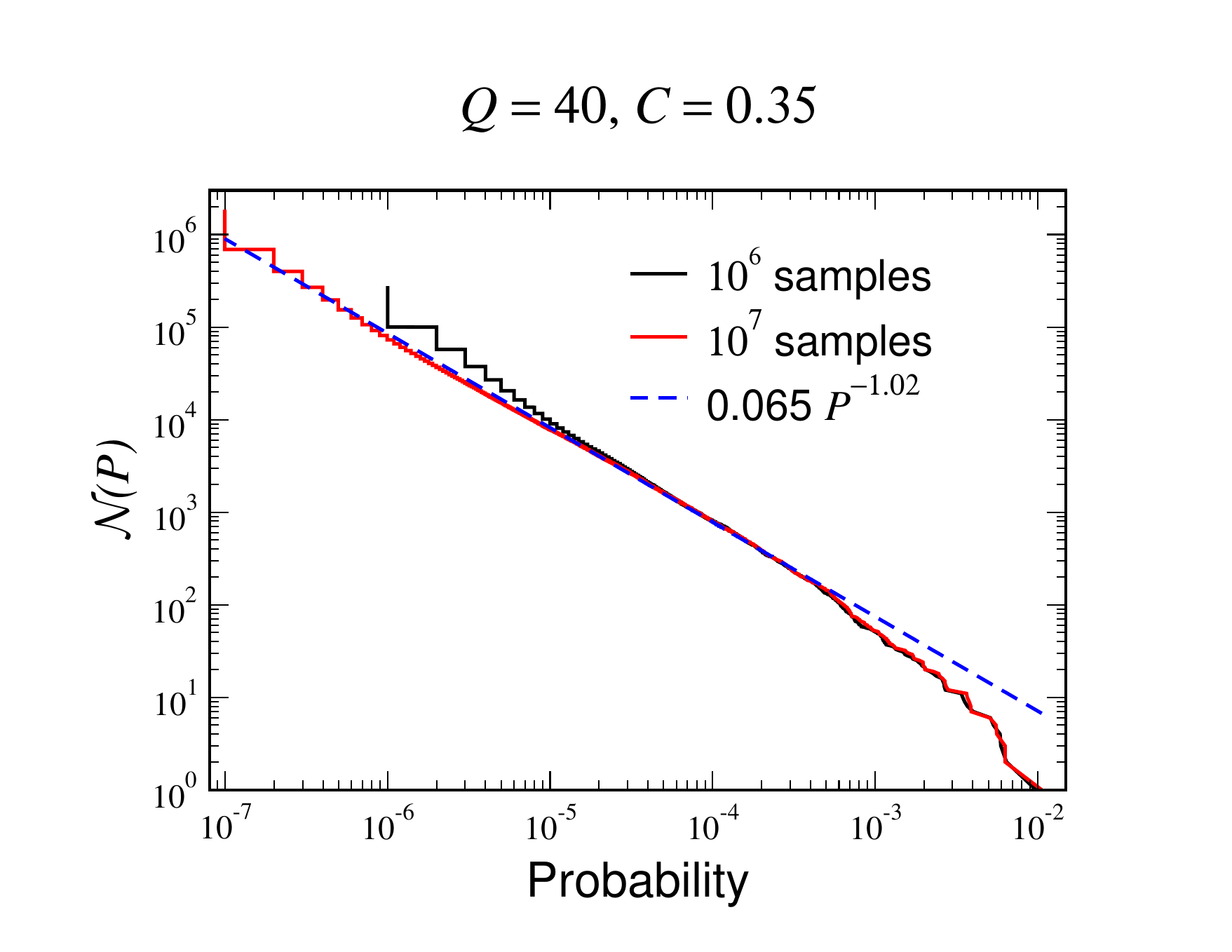}}
 \caption{The complementary cumulative distribution function ${\cal{N}}(p)$; i.e., the number of states with probability greater than $p$, for runs of $10^6$ or $10^7$ consecutive snapshots, presented in log-log scale. $Q=40$, $K=200$, $C=0.35$,
 $\sigma=1/\sqrt{8}$, $\lambda=0.01$. The time between snapshots is  $0.08/\lambda$.  Also shown is a power-law fit to the low-$p$ data.}
 \label{figC35d}
 \end{figure}

 In order to meaningfully measure the entropy, we need to reduce the dimensionality of the state space.  Essentially, we wish to classify a state by which species are stably present.  To do this it is necessary  to eliminate the very low abundance species, with only 1 or 2 representatives, since their presence is in general very transitory.  A convenient way to accomplish this without introducing an arbitrary threshold  is via the instantaneous ``quenching" procedure described in Ref. \cite{kessler1}.  Each snapshot is taken as the initial condition of a deterministic run with $\lambda=0$, with equations of motion 
 $\dot{n}_i = (1 - d_i)n_i ,$
 which is run until it reaches a steady-state.  In this way, low abundance species with negative net growth rates are eliminated.  We then calculate the number of appearances of every state in a long run, which when normalized by the number of snapshots gives the probability $p$ of that state. From this, we generate the complementary cumulative distribution function, ${\cal{N}}(p)$, the number of states whose probability is greater than a given $p$.  The results of a typical run for $Q=40$ is given in Fig. \ref{figC35d}.  We see that ${\cal{N}}(p)$ behaves as a power-law for all but the largest $p$'s, over a span of 4.5 decades.  This translates directly to an asymptotically linear $S(E)$ relation as $E \equiv -\log {p} \to +\infty$. As discussed in detail in Ref.~\cite{bialek2} , this means that the attempt to find a typical value of $E$ at the fixed temperature $T=1$ via the thermodynamic relation $\frac{dS}{dE} =1$ cannot succeed, and the system exhibits a rather unusual type of critical behavior. 

\begin{figure}
 \includegraphics[width=0.28\textwidth]{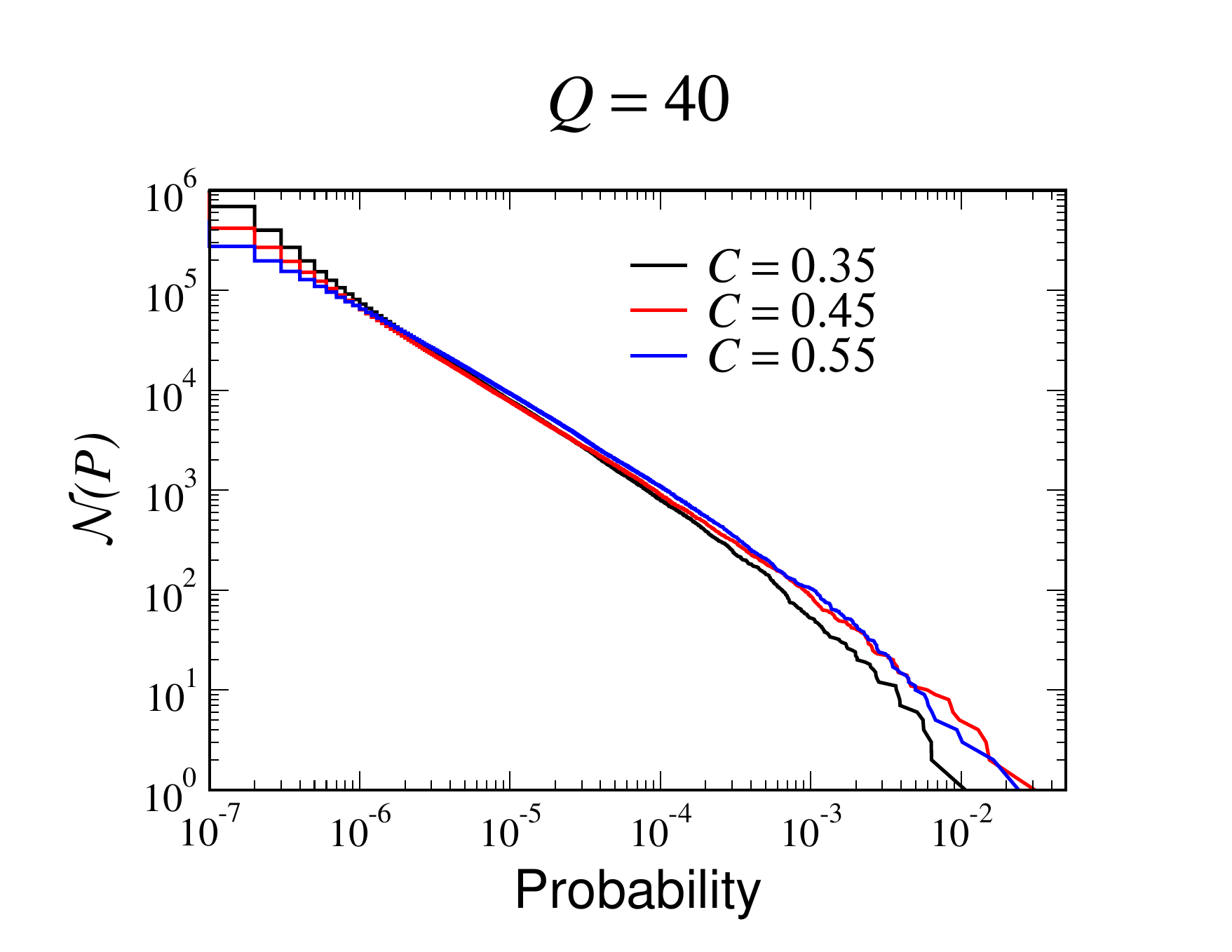}\hspace*{-0.3in} \includegraphics[width=0.28\textwidth]{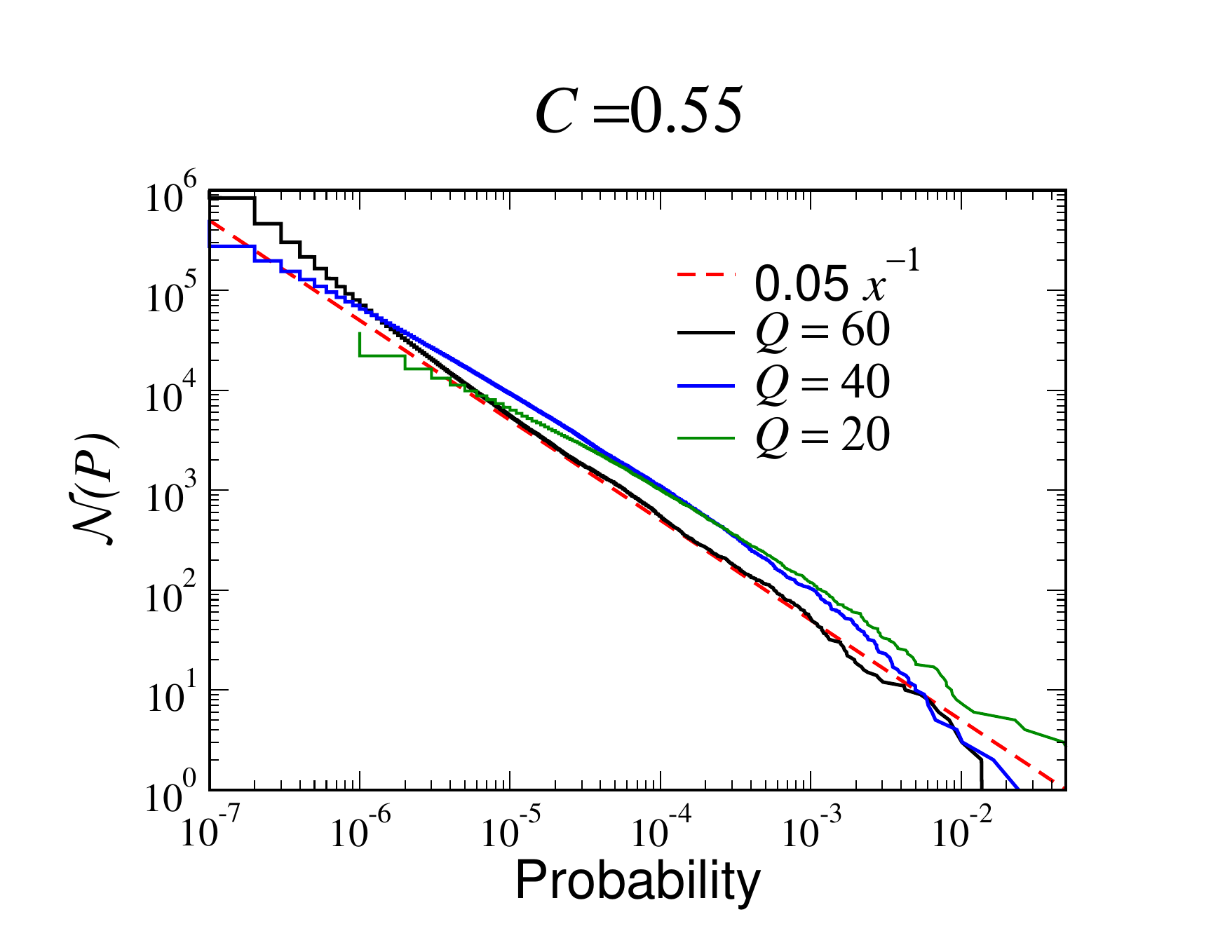}
 \caption{The complementary cumulative distribution function ${\cal{N}}(p)$; i.e., the number of states with probability greater than $p$, for $C=0.25, 0.45, 0.55$  with $Q=60$ in (a) and $Q=20,40,60$ with $C=0.55$ in (b). The other parameters are $\lambda=0.01$, $K=50Q$, $\sigma=.5\sqrt{20/Q}$. The time between snapshots is  $0.08/\lambda$.  }
 \label{figC0manyC}
 \end{figure}

\begin{figure}
 \includegraphics[width=0.45\textwidth]{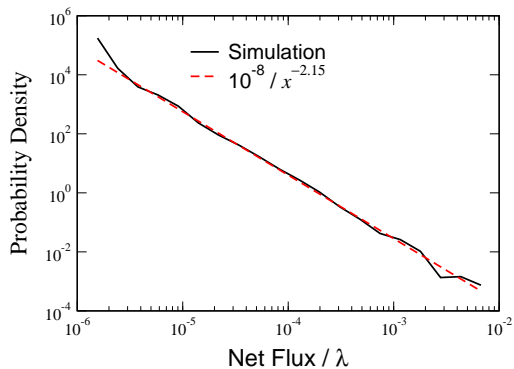}
 \caption{Distribution of net fluxes (in units of the immigration rate $\lambda$) for the case $Q=20$, $C=0.45$, $\sigma=0.5$, $\lambda=0.01$.}
 \label{figflux}
 \end{figure}

 One question that immediately arises is how general this phenomenon is, or whether it holds only for a specific set of parameters or range of parameters.  To investigate this, we varied $C$ in the range from $0.35$ to $0.55$, measuring ${\cal{N}}(p)$ for each.  We also varied $Q$, the number of interacting species. As $Q$ is changed, we scale the variance parameter $\sigma$ of the inhibition matrix as $1/\sqrt{Q}$~\cite{may} and $K$ linearly with $Q$. The results are presented in Fig. \ref{figC0manyC}.
 Of interest is the fact that the linearity improves as $Q$ increases and that the data approaches a pure Zipf law~\cite{newman,nemenman} with scaling exponent unity.
 The fact that ``criticality" extends for a finite parameter range in $C$ is presumably due to the inherently non-equilibrium nature of the problem. One way to measure deviations from equilibrium is via the violation of detailed balance in the flux between states. In Fig. \ref{figflux}, we present a plot of the net steady-state flux between different states in the model, showing that these again have a power-law distribution. Of course these would all be precisely zero in any model where the steady-state obeyed detailed balance.  
 
 {\bf Neural Net Model} The main nonlinear ingredient of our ecology model is the dynamic competition between different species. We are thus motivated to turn to a model with a similar interaction in a different context, a set of neurons which are mutually inhibitory and are driven by external sources~\cite{brunel}. The inhibition matrix $J_{i,j}$ is constructed from random elements of average size unity (distributed uniformly between 0 and 2), with density $\rho$, with all other elements set to zero. As in standard integrate and fire networks, the voltage level on a given neuron decays exponentially with time constant $\tau$.  The external input, $V_i^I$ is uniformly distributed at each time slice $dt$ between $dt(1\pm \sqrt{3} \sigma_I)$.  The threshold for a neuron firing is given initially by
\begin{equation}
V_c = \tau \left(1 - e^{-T_p/\tau}\right)
\end{equation}
so that in the absence of noise, every neuron would fire with period $T_p$.  Thus, the update rule for $V_i$ is
\begin{equation}
V_i(t+dt) = e^{-dt/\tau}V + V_i^I
\end{equation}
At this point, suprathreshold neurons fire, and are reset to 0. The firing neurons then reduce the voltage on the neurons they are connected to according to
\begin{equation}
V_i = V_i - C\sum_{j\in \textit{fire}} J_{i,j} 
\end{equation}
A record is kept of every neuron that fires during each window of duration $T_W$, which in this case is reasonable to take as $T_p$, since this is the minimum interval between firings (modulo the input noise).

\begin{figure} \vspace{.2cm}
 \includegraphics[width=0.225\textwidth]{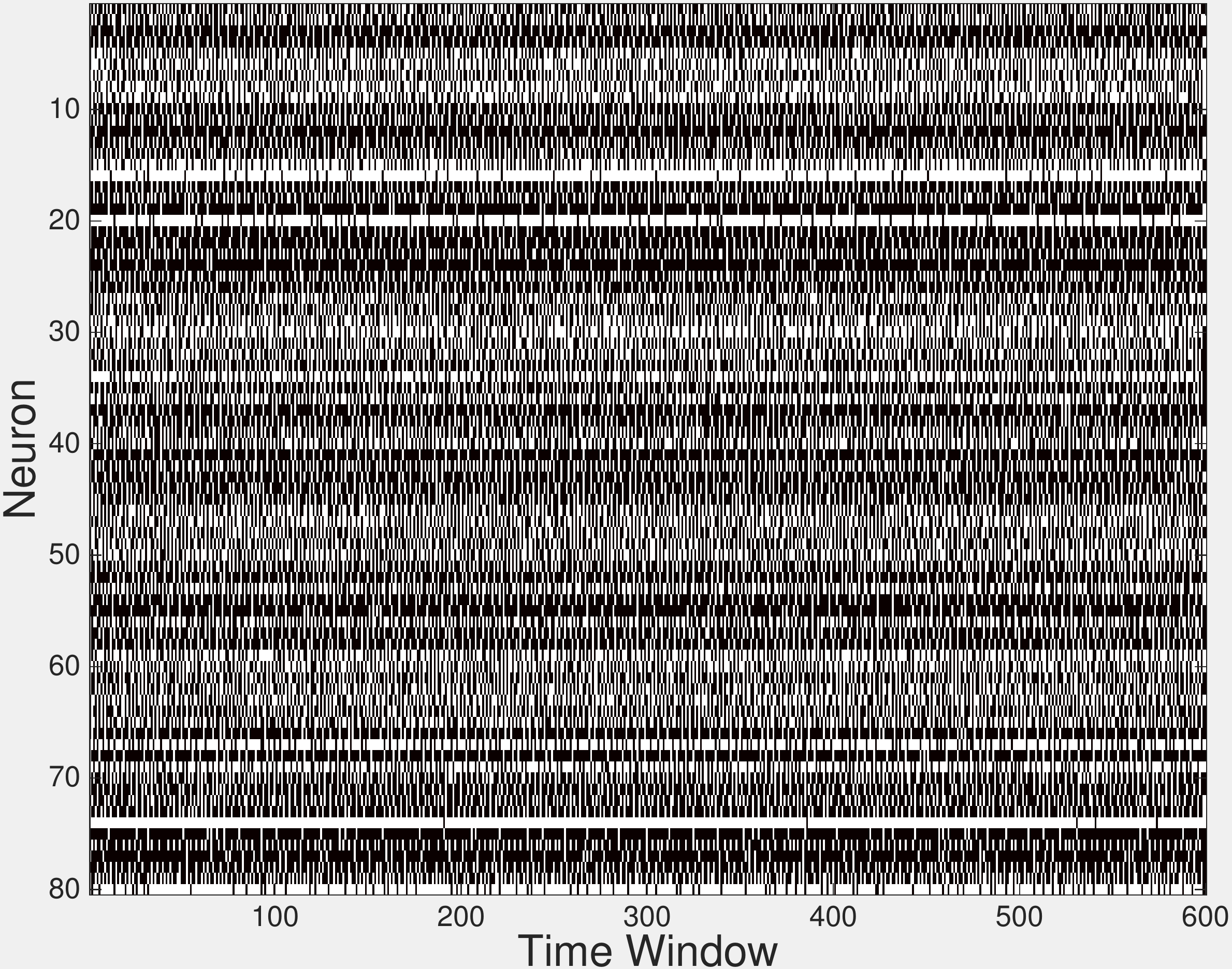} \includegraphics[width=0.225\textwidth]{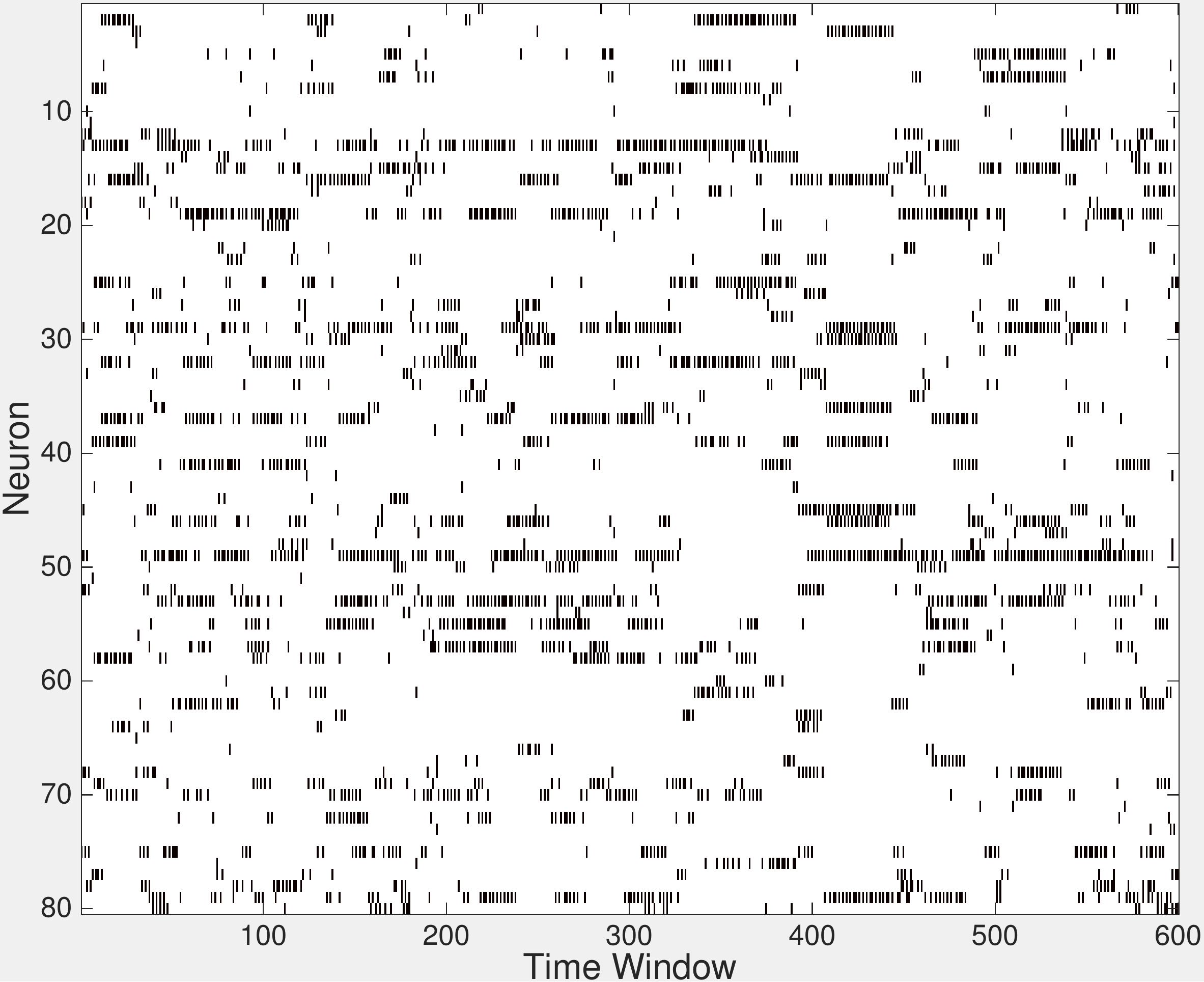}
 \caption{Sample firing patterns from our neural net model with $N=80$ neurons. Left: Pattern with constant threshold $V_c = 19.67$. The time window was $T_W=T_p=83$. Right: Pattern with dynamical thresholds, governed by the parameters $f_T=0.85$, $\gamma_T=0.02$.  Due to the lower threshold of the adapted neurons, the time window was taken as $T_W=40$. For both models, $C=1$, $\sigma=0.7$, $\sigma_I=0.02$, $\tau=20$.}
 \label{fig0}
 \end{figure}

A pictorial view of the dynamics is given in Fig. \ref{fig0}a. Notice that the dynamical state of the system is very different than that depicted in Fig 1c, which presents a similar trace for the colony model in the range of parameters we have been considering. Essentially, the strong inhibition and the lack of any long-term memory that a neuron has been active serves to create an extremely chaotic process; in the ecology model the role of memory is played by the size of a species population which if large prevents the species from rapidly disappearing. To mimic this effect, we added a feedback between firing and threshold. When a neuron fires, its threshold for future firing is lowered, by a factor $f_T$. The threshold then relaxes exponentially back to its nominal value, $V_c$, with a relaxation rate of $\gamma_T$. One could imagine accomplishing analogous changes by altering synapses, i.e. weakening the synaptic inhibition for any neuron that fires. Once this change is implemented, the dynamics stabilizes to the form seen in Fig. \ref{fig0}b. Somewhat remarkably, the actual variation in thresholds achieved by our algorithm once the system has settled into a statistical steady-state is only about 10\% (data not shown). Yet, this is enough to create robust patterns of firing and non-firing neurons which are then acted upon by the fluctuation input signals, since the period of firing of a neuron is exponentially sensitive to its threshold when the threshold is close to its critical value of $\tau$. This capability of using relatively weak induced memory to create stable firing patterns should be studied in more detail in future works. 

\begin{figure}
\includegraphics[width=0.45\textwidth]{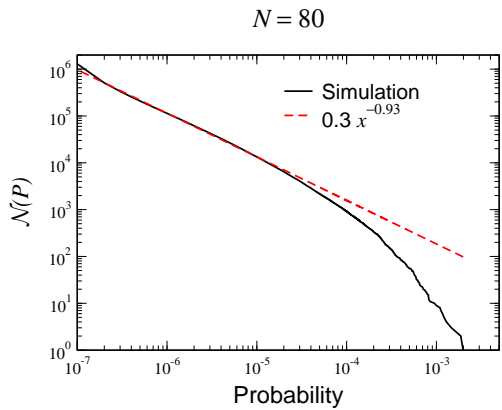}
\caption{Complementary cumulative distribution function for $N=80$. Parameters are the same as in the simulation shown on the right side of Fig. \ref{fig0}}
\label{fig1}
\end{figure}

We then tabulate the statistics of firing patterns.  Our first set of runs is with $N=80$ neurons and parameters  $\rho=1$, $\tau=20$, $T_p=83$, $C=1$, $\sigma_I=0.02$.  In Fig. \ref{fig1}, we again turn to the complementary cumulative distribution function. As in the previous model, we now observe an almost perfect Zipf law form~\cite{zipf}, scaling with exponent close to -1. This would guarantee that any attempt to model the thermodynamics of the system with an equilibrium Hamiltonian would ``discover" that the system looks like it had been tuned to be very close to a critical point. Again however, this state exists over a broad range of parameters and does not require any fine-tuning. A similar tabulation for the non-adaptive model shows that the system thermodynamics is completely dominated by a huge number of very low occupancy states.  Approaches which claim that scaling should exist in most models of this kind \cite{nemenman} need to be able to explain what fails in our non-adaptive neural system.

Scaling behavior of various kinds seems to be a ubiquitous feature of many non-equilibrium systems~\cite{soc}. We have seen here two examples of such systems. We are certainly not claiming that all or even most models one could write down will automatically have such a non-trivial property. As shown in our neural example, we need to have the proper mix of memory and transitions; too little memory and the system has no enduring high probability states and too much memory (for example, by lowering the inhibition) locks the system permanently into too few states. But unlike the expectation for equilibrium distributions, there is a "Goldilocks" range, not a "Goldilocks" point. This of course would make it much easier for evolution to drive biological processes to have these properties, should they prove advantageous for information processing in response to stimuli.

\begin{acknowledgments}
This work was partially supported by the National Science Foundation (HL) through the Center for Theoretical Biological Physics, PHY-1400968 and the Israel Science Foundation (DAK) 376/12.

\end{acknowledgments}

\bibliographystyle{pnas}
\bibliography{scaling}
\end{document}